# Title: Electronic excitation of ultrafast collective amorphous-amorphous transitions in glassy phase-change material

**Authors:** Yingpeng Qi[1,3*], Nianke Chen[2*], Zhihui Zhou[3], Qing Xu[3], Yang Lv[3], Xiao Zou[3], Tao Jiang[3], Pengfei Zhu[3], Min Zhu[4], Dongxue Chen[1,5], Zhenrong Sun[6], Xianbin Li[2*], Dao Xiang[3,7*]

**Affiliations:**

[1]Suzhou Institute for Advanced Research, University of Science and Technology of China, Suzhou 215123, China

[2]State Key Laboratory of Integrated Optoelectronics, College of Electronic Science and Engineering, Jilin University, Changchun 130012, China;

[3]Key Laboratory for Laser Plasmas (Ministry of Education), School of Physics and Astronomy, Shanghai Jiao Tong University, Shanghai 200240, China.

[4]School of Integrated Circuits, Shanghai Jiao Tong University, Shanghai 200240, China.

[5]Department of Physics, University of Science and Technology of China, Hefei, Anhui 230026, China

[6]State Key Laboratory of Precision Spectroscopy, and School of Physics and Electronic Science, East China Normal University, Shanghai 200241, China

[7]Zhangjiang Institute for Advanced Study, Shanghai Jiao Tong University, Shanghai 201210, China.

*Correspondence to qypcool@ustc.edu.cn, chennianke@jlu.edu.cn, lixianbin@jlu.edu.cn, dxiang@sjtu.edu.cn.

**Abstract:** The intrinsic nature of glass states and glass transitions at the atomic scale remain a fundamental open question in condensed-matter physics and materials science. The key to solving the glass transition problem lies ultimately in achieving a complete understanding of the physics governing the structural relaxation and its dynamic properties. Nonetheless, directly probing dynamic atomic-scale structural changes in order to identify the precise local structural motifs and establish quantitative structure-property relationships remains an outstanding challenge. By combining femtosecond electron diffraction with time-dependent density-functional theory molecular dynamics simulations, we directly capture ultrafast amorphous-amorphous transitions indicated by collective bond stretching (< 0.2 ps) and angle bending (~ 0.5-2 ps) in glassy phase-change material GeTe. The ultrafast bond stretching is accompanied by localized oscillation modes with the frequency of ~ 3.10 THz, unambiguously signaling the local Peierls-like bonding structure and the flexibility of these polarized bonds. These ultrafast collective atomic motions, captured across timescales ranging from femtoseconds to picoseconds, directly reveals the structural origin of the boson peak and provide compelling evidence for many-body interactions in amorphous materials. Furthermore, the ultrafast amorphous-amorphous transitions induce a drastic insulator-metal transition, directly revealing both the underlying switching mechanism and the fundamental speed limit of the ovonic threshold switch. These insights establish a fundamental framework for rationally engineering relaxation pathways and phase-change/threshold switch in amorphous materials. Femtosecond electron diffraction provides a powerful novel approach to deciphering the structural complexity and functional mechanisms of amorphous materials by resolving collective atomic motions from random diffusion dynamics in the time domain.

**Main Text:** Glasses are ubiquitous in nature and have a wide range of applications in our daily lives as well as in modern technology due to the unique physical, chemical and mechanical properties distinct from the crystal counterparts [1,2]. As a vitreous supercooled liquid, the glassy material is in a thermodynamically metastable state between the molten liquid state and the crystalline state, featured with many-body interactions and strong disorder. The intrinsic nature of glass states and glass transitions remains a major unsolved problem in condensed matter physics and materials science [3-8]. The Zachariasen paradigm [9] posits that the short-range structure (≤ 5 Å) in glasses closely resembles the atomic arrangement of their crystalline counterparts and the random nature of the inter-unit connections produces disorder at long length scale. This paradigm is also known as the continuous random network model. However, referring to specific material systems, this simple model is not without challenge [10-15] and the spatial correlation beyond the first nearest neighbor remarkably contributes to the structure–property relationships in glasses [16-18]. Besides the inherent structure, the fundamental understanding of the diverse properties and complex phenomena in glasses, such as the medium range order (5-20 Å) [11,19,20], the boson peak [21-24], the fast β relaxation [25-27] and the formation of a nucleus and crystal growth [13,28,29], also remain largely fragmented. The many years of researches lead to a consensus that the structural dynamics or structural relaxations, based on the concept of potential energy surfaces (PES) and collective atomic motions [30,31], serve as the link between the intrinsic structure and the complex phenomena in glasses [32-34]. Therefore, beyond conventional spectroscopic and scattering measurements [19,35-40] and theoretical simulations [41-44] in an equilibrium state, resolving the collective structural dynamics corresponding to PES with ultrahigh spatiotemporal resolution lays the foundation of a unified framework to describe the structures and the properties in amorphous materials. It is noteworthy that the strong excessive vibrational modes in the terahertz range (~ 1-10 meV) of the boson peak [21-24] and the atomic-scale structure adjustments over a few bond distance in the β relaxation [27,42,45-47] call for femtosecond and sub-angstrom spatiotemporal resolution. To the best of our knowledge, femtosecond electron\x-ay diffraction [48-51] is the exclusively appropriate approach, while it has rarely been explored in the field of glassy materials because the complexity behind the disordered structure has long been overlooked.

In this work, we take the amorphous phase-change material GeTe as a model system because it has the typical characteristics and properties of glassy systems with a simpler binary composition. GeTe has been effectively exploited in non-volatile memory and neuromorphic computing devices [52] due to the strong optical/electrical contrast between crystal and amorphous phases, however, an intimate understanding of the amorphous structure and the phase transition on the atomic-scale level is still not clear [31]. The crystalline phase of GeTe is rhombohedral, featuring a Peierls distortion that manifests as alternating short and long bond on opposite sides of the central atom [53,54]. In contrast, several competing models have been proposed to describe the intrinsic atomic arrangements in amorphous GeTe, including the local tetrahedral model [10,55], the local octahedral model incorporating Peierls-like distortions [25,56,57] and other conceptional models [13], with extensive investigations using X-ray absorption spectrum, scattering techniques and advanced numerical simulations. The bonding change-whether attribute to the unique metavalent bonding [14] or to the modulated polarizability [15]-between the crystalline and the amorphous phase continues to spark debate. The physical mechanism of boson peak in phase-change materials [58] and other glassy materials-whether stemming from disordered atomic packing [21] or from spatial fluctuations in the interatomic forces constants [59,60]-remains unresolved. During the structural switching process, the crystallization process is the time-limiting step. The crystallization speed has been improved from tens of nanoseconds to hundreds of picoseconds by introducing prestructural ordering [61] and crystal precursors [62]. However, the primary timescale governing nucleation and crystallization and the ultimate speed limit of these processes remain experimentally unestablished in phase-change materials. When subjected to an electric field, certain amorphous materials undergo a transition from a high resistance state to a conduction state, a phenomenon known as ovonic threshold switching. This effect has recently spurred a major technological breakthrough in the memories field [63-65]. However, the physical origin of the threshold switching above a critical electric field-whether it arises from atomic relaxation or electronic relaxation, remains a subject of active debate [63-65].

In this study, we tackle the aforementioned universal challenges in amorphous materials by probing the ultrafast collective atomic motions in amorphous GeTe using femtosecond electron diffraction. Figure 1A and 1B illustrate our approach and key findings. The ~50 fs temporal resolution of our femtosecond electron diffraction system [66] allows for the detection of the fastest collective motions in crystalline disordered materials [48,50] and amorphous materials with picometer-scale spatial resolution. With this novel approach, we reveal ultrafast amorphous-amorphous transitions indicated by bond stretching (< 0.2 ps) and angle bending (~ 0.5-2 ps), which are well reproduced by the time-dependent density-functional theory molecular dynamic. As illustrated in Figure 1B, the ultrafast bond stretching gives rise to an intermediate state and the subsequent angle bending relaxes the local structure to a more stable basin of the PES. The overall structural evolution signifies an incubation process toward the crystalline phase. Based on these ultrafast collective atomic motions, we construct a unified scenario of the intrinsic amorphous structure and the fundamental properties in amorphous materials. First, the ultrafast Ge-Te (Ge) bond stretching and the localized oscillators (~ 3.10 THz) unambiguously identify the local Peierls-like bonding structure and the flexibility of these local bonds. Second, the femtosecond to picosecond structural responses together with the multiple local minimums of the potential energy landscapes expanding ~ 1.2 Å, elucidate that the excessive vibrational modes of the boson peak derives from the random fluctuation of the forces constants. The vibrational modes of the defective Ge-Ge and Ge-Ge-Ge (two dimensional) homopolar structural motifs, which are absent in the crystalline phase, suggest the disordered atomic packing as an alternative origin of the excessive vibrational modes [21]. Third, the ultrafast electronic delocalization by bond stretching and the reduction of the Ge-Ge wrong bonds signify an ultrafast suppression of configurational entropy and initiate an ultrafast incubation process for nucleation, against the physical aging (β relaxation) [25,27]. Fourth, the ultrafast amorphous-amorphous transitions triggered by electronic excitation induce a drastic insulator-metal transition, directly demonstrating the underlying mechanism of the ovonic threshold switch.

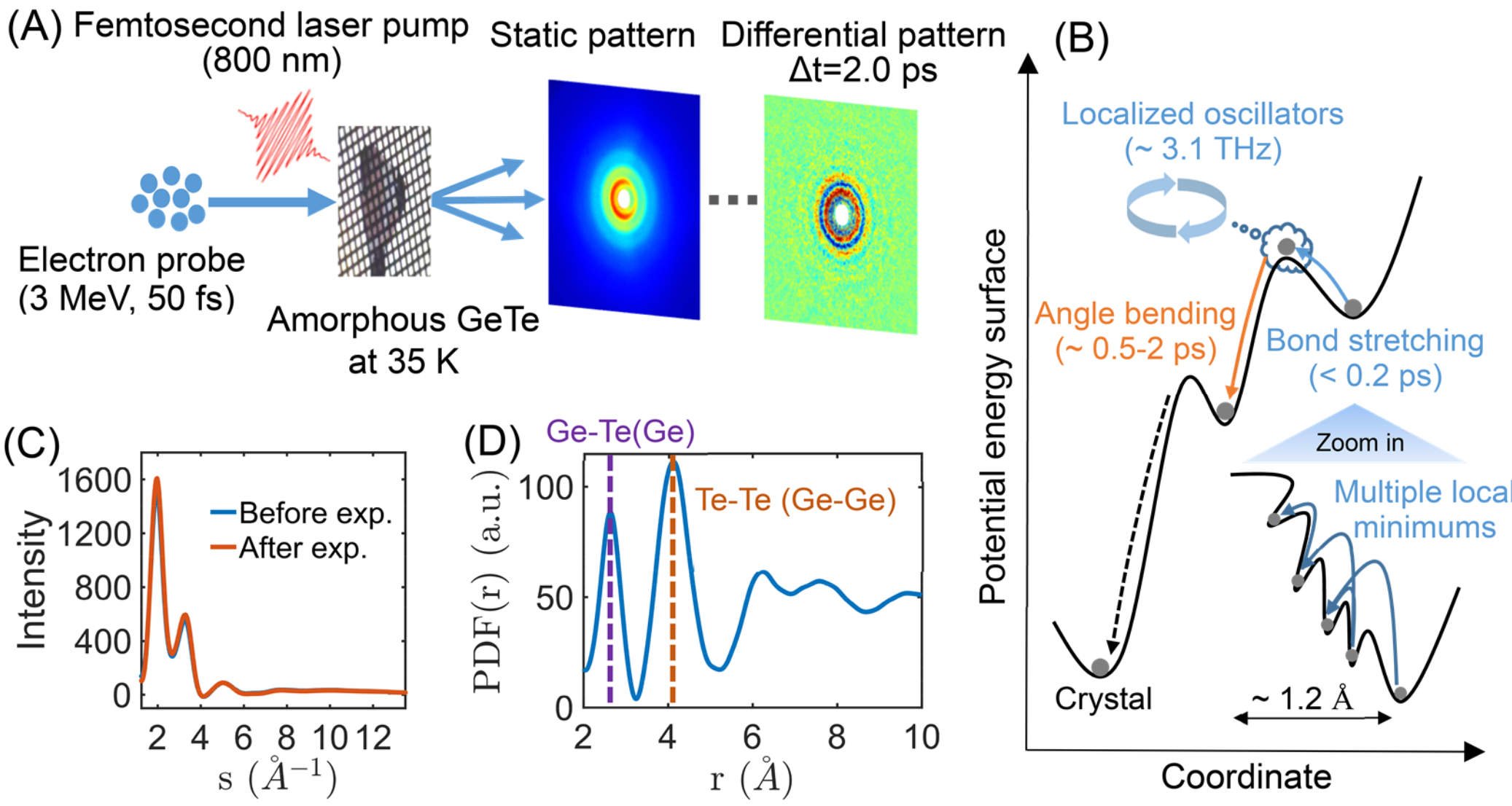

**Fig. 1. Femtosecond electron diffraction reveals ultrafast amorphous-amorphous transitions in GeTe. (A)** Schematic representation of the femtosecond electron diffraction system. The static diffraction pattern of the amorphous GeTe and the differential pattern at the time delay of 2.0 ps are displayed. **(B)** Schematic representation of the photoinduced amorphous-amorphous transitions, i.e. the bond stretching and angle bending, within the framework of potential energy surface changes. The localized oscillators (~ 3.10 THz) and the multiple local minimums over the ~ 1.2 Å range accompanying the bond stretching are revealed and illustrated. The dotted black line suggests the further relaxation towards the crystalline state. **(C)** Radial averaged diffraction intensity before and after experiments. **(D)** Reduced pair distribution function (PDF) G(r) of amorphous GeTe with main peaks of Ge-Te (Ge) and Te-Te (Ge-Ge) bonds.

## The ultrafast amorphous-amorphous transitions

The amorphous GeTe with the thickness of 20 nm is deposited on the NaCl substrate and subsequently transferred onto a TEM copper grid for ultrafast electron diffraction experiments. Fig. 1A schematically illustrates the femtosecond electron diffraction setup (see details in Supplementary Materials). The sample is excited with the 800 nm femtosecond laser (above band-gap excitation) at the base temperature of 35 K. This ultralow base temperature is essential to resolve the boson peak feature, which typically manifests most prominently in the 10-30 K range. By varying the time delay between the pump laser and the ultrashort electron probe pulse, we capture time-resolved intensity changes of the diffraction patterns, enabling direct observation of atomic-scale vibrational and displacive dynamics. To accurately characterize the intrinsic structural feature, the pump fluence is carefully calibrated to prevent laser pumping induced irreversible relaxation. As shown in Fig. 1C, the radial-averaged diffraction intensities recorded before and after the experiment are nearly identical, confirming the structural integrity and stability of the amorphous phase throughout the measurements. Furthermore, the experimentally obtained pair distribution function (PDF), displayed in Fig. 1D, exhibit excellent agreement with previous reported PDF for amorphous GeTe [13,57], attesting to the high quality of the as-prepared amorphous film. The first two prominent peaks at ~ 2.65 Å and ~ 4.12 Å correspond to the Ge-Te (Ge) and the Te-Te (Ge-Ge) correlations, respectively. The pair distribution function (PDF) G(r) as a function of distance at the time delay of -0.5 ps and -1.5 ps is displayed in Fig. S1 in Supplementary Materials.

Upon femtosecond laser excitation, significant structural dynamics are observed both in reciprocal space (Fig. 2A-2B) and real space (Fig. 2C-2D). The Fourier transformation from reciprocal to real space and the calculation of the PDF change,

ΔPDF [67], are detailed in Supplementary Materials. In reciprocal space, the percent change of the radially averaged diffraction intensity exhibits two key features across the three scattering regions shown in Fig. 2B: an ultrafast intensity change within the first 0.2 ps and an antiphase damped oscillation of the percent difference (PD) signal marked by the solid vertical line. In real space, this same oscillation manifests more distinctly as shown in Fig. 2D, with a center frequency of ~ 3.10 THz. This coherent oscillation indicates a localized oscillator with the frequency aligning with the 3.33 THz coherent phonon previously observed in crystalline GeTe, which arises from ultrafast suppression of the Peierls distortion [53]. Therefore, the localized vibrational mode unambiguously evidences the local Peierls-like bonding structure featured with electronic polarization of Ge atom between the two bonds to the left and right, akin to the crystalline phase [68]. Besides the vibrational mode, the ultrafast intensity decrease (increase) within 0.2 ps in the 2.3-2.6 Å (2.8-3.5 Å) region suggests the stretching of the Ge-Te (Ge) bond. Consistently, the first peak in the PDF (Fig. 2E) shows the signature of the center position change to a further distance. Quantification of this shift (Fig. 2F) reveals a rapid, sub-0.2 ps displacement as well a highly damped oscillation-directly corroborating both the ultrafast bond stretching and the coherent oscillatory component in Fig. 2D. We attribute this ultrafast structural response to the delocalization of the bonding electrons upon photoexcitation, followed by Coulomb-driven coupling between the excited electrons and and the nuclei-prompting bond stretching and launching the local oscillators. Notably, the intensity evolution in Fig. 2F proceeds much slower than the positional shift, suggesting the involvement of additional, slow relaxation channels, such as the angle bending mode discussed further in the next section.

With a slight higher pump fluence, the ultrafast bond stretching and the localized oscillations are reproduced, as shown in Fig. S2 in Supplementary Materials. The oscillatory feature in Fig. 2D spans a broad range of ~ 2.3-3.5 Å in real space, significantly broader than the sharply defined bond lengths typically seen in crystalline counterparts. Thermal vibrations induced expansion is strongly suppressed because the ultralow base temperature of ~ 35 K employed in the experiment. Therefore, this wide oscillation envelope reflects a quasi-continuous distribution of Ge-Te (Ge) bond lengths centered around 2.3-3.5 Å, which induces substantial spatial fluctuations in the local forces constants and gives rise to pronounced excess vibrational modes peaked near ~ 3.10 THz. This quasi-continuous bond length distribution corresponds to multiple shallow local minima in PES, as illustrated in Fig. 1B-evidence of strong many-body interactions. As a result, the ultrafast bond stretching proceeds via multiple transitions among adjacent local minima. Such a picture-where the random fluctuation of the forces constants drives excess vibrational modes-is fully consistent with the theoretical framework of the boson peak [59,60]. Importantly, this ultrafast bond stretching dynamics suppresses the electronic localization along the Ge-Te bonds and reduces the population of defective Ge-Ge wrong bonds, thereby promoting the subsequent nucleation and crystallization processes.

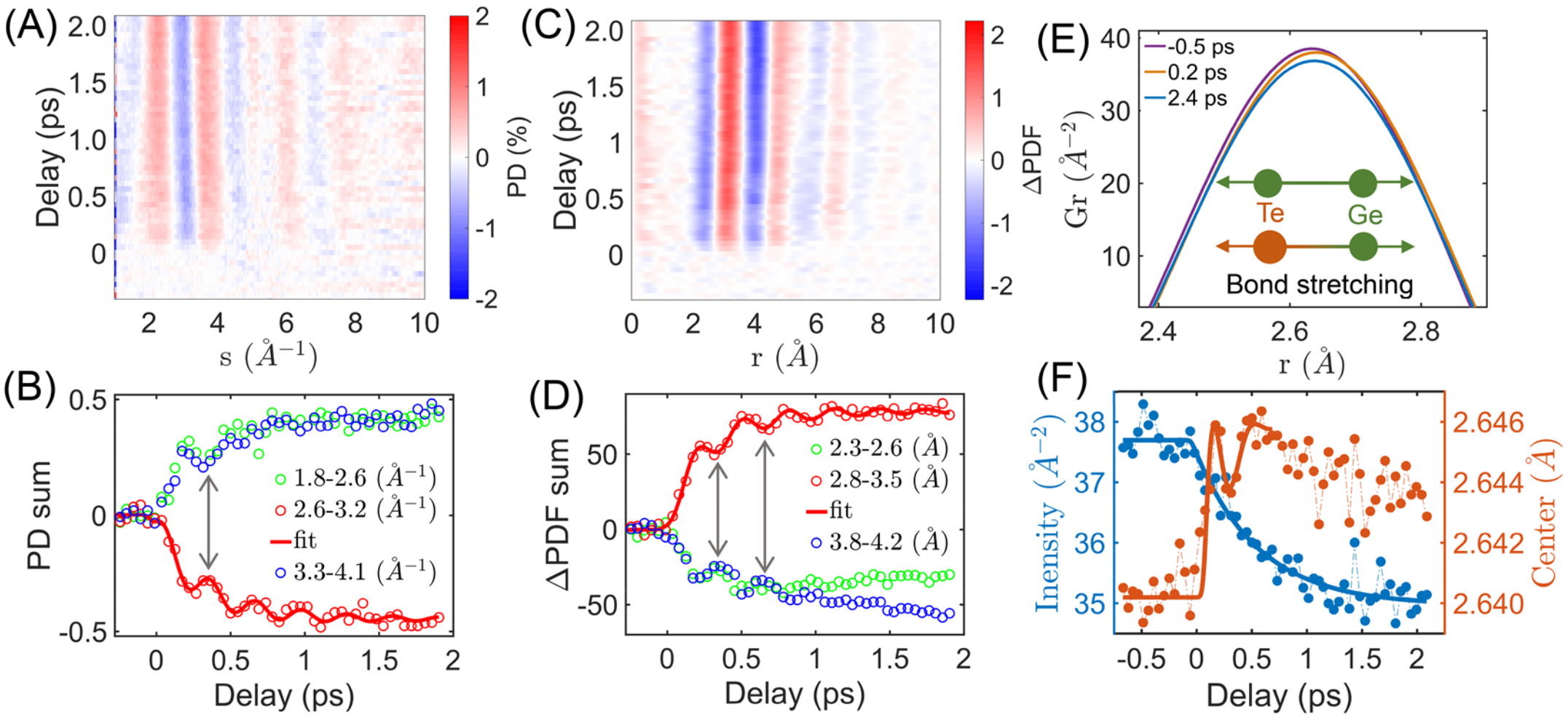

**Fig. 2. Ultrafast oscillation and stretching of the local Ge-Te (Ge) bond in the reciprocal (A-B) space and the real space (C-D) with the pump fluence of 2.0 mJ/cm$^2$. (A)** In reciprocal space, the percent difference (PD) signal of the radially averaged diffraction intensity in a false colour map. **(B)** PD signal across three scattering regions as a function of time delay. **(C)** In real space, ΔPDF in a false colour map. **(D)** Temporal evolution of the ΔPDF at three distinct distances. The solid red line is the fit with a combination of an exponential function and a cosine function. The vertical solid lines indicate the antiphase oscillations occurring at different distances. **(E)** Intensity distribution of the first peak in real space at three time delay points. The inset is the schematic illustration of the Ge-Te (Ge) bond stretching. **(F)** Time domain evolution of the intensity and the position of the first peak in real space. The solid lines are fits to the experimental results.

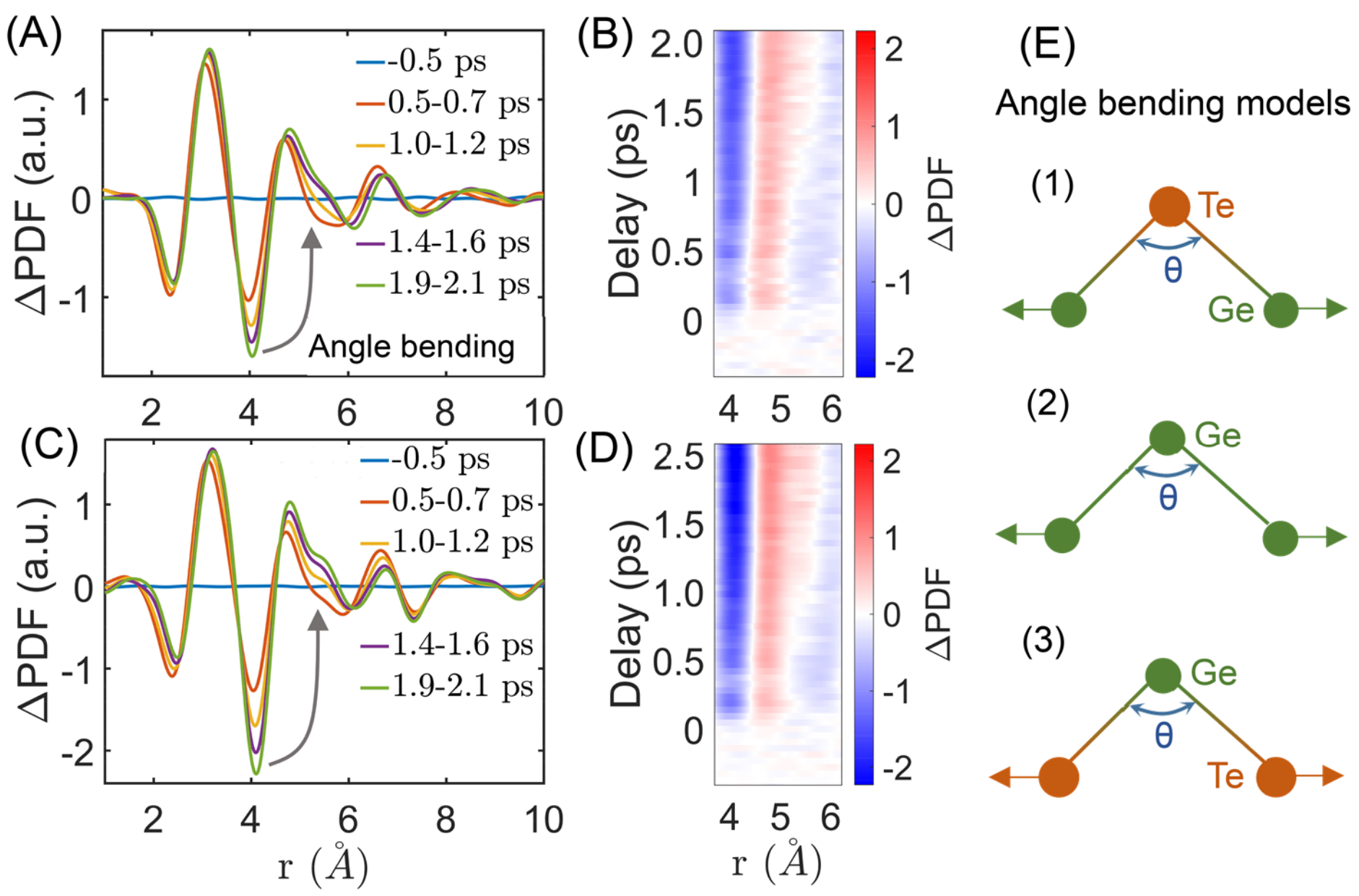

**Fig. 3. Ultrafast angle bending of the local structural motifs. (A)** ΔPDF as a function of distance at varying time delays. **(B)** ΔPDF in a false colour map. The pump fluence in **(A)** and **(B)** is 2.0 mJ/cm$^2$. The solid arrow indicates the anticorrelated change of ΔPDF around ~ 4.0 Å and ~ 5.4 Å. **(C)** ΔPDF as a function of distance at varying time delays. **(D)** ΔPDF in a false colour map. The pump fluence in **(C)** and **(D)** is 3.0 mJ/cm$^2$. **(E)** Schematic presentation of the three angle bending models.

Except for the ultrafast Ge-Te (Ge) bond stretching within ~ 0.2 ps, the intensity in the 3.8-4.2 Å range-corresponding to the Te-Te (Ge-Ge) distance-decays continuously over the subsequent 0.5-2 ps, as shown in Fig. 2C-2D and Fig. S2 in Supplementary Materials. To quantify this evolution, the ΔPDF as a function of distance at varying time delays is shown in Fig. 3A. Beyond 0.5 ps, the intensity of ΔPDF around 4.0 Å decays further, while that near 5.4 Å undergoes a pronounced sign reversal-from negative to positive. Aside from these two distinct features, the ΔPDF remains largely unchanged across other regions within 0.5-2 ps. This crosover is vividly visualized in the false colour ΔPDF map in Fig. 3B, which highlights the emergence of positive intensity around 5.4 Å with increasing time delay. As increasing the pump fluence to 3.0 mJ/cm$^2$, all aforementioned features persist, as shown in Fig. 3C-3D. The anticorrelated intensity changes at ~ 4.0 Å and ~ 5.4 Å strongly suggest a redistribution of the local structural motif: a shift in the distance distribution from shorter ~ 4.0 Å to longer ~ 5.4 Å. In real space, the distance distribution around ~ 4.0 Å arises predominantly the Ge-Ge distance and the Te-Te distance, both of which are associated with threefold coordinated triangular motifs featuring bond angles close to 90º [13], as illustrated in Fig. 3E. The observed anticorrelation thus points to an increase in the Ge-Ge and/or Te-Te distance by angle bending. Such angle bending motions with three particles evidence the lowest-order many-body correlated interaction involving angular information. Among the three possible angle bending configurations depicted in Fig. 3E, the first two, i.e. the Ge-Te (Ge)-Ge motif, is more plausible: the heavier Te atom acts as a relatively immobile hige, facilitating preferential displacement of the lighter Ge atoms. Furthermore, given the prevalence of Ge-Ge homopolar bonds in the amorphous phase of GeTe, the Ge-Ge-Ge motif is expected to contribute significnatly to the angle bending.

As shown in Fig. 3A-3D, the angle bending process begins at ~ 0.5 ps and reaches quasi-equilibrium at ~ 2 ps. This timescale corresponds to a vibrational property with the frequency of 0.5-2 THz, which is well within the range of the vibrational modes of the boson peak. The Ge-Ge-Ge motif with homopolar bonds is uniquely present in the amorphous phase and entirely absent in the crystalline phase, therefore, the observed ultrafast angle bending dynamics constitute the first compelling evidence that defective local structures contribute to the excessive vibrational modes underlying the boson peak [21]. Moreover, the ultrafast angle bending indicates the dissociation of the homopolar bonds, thereby promoting the subsequent nucleation and crystallization.

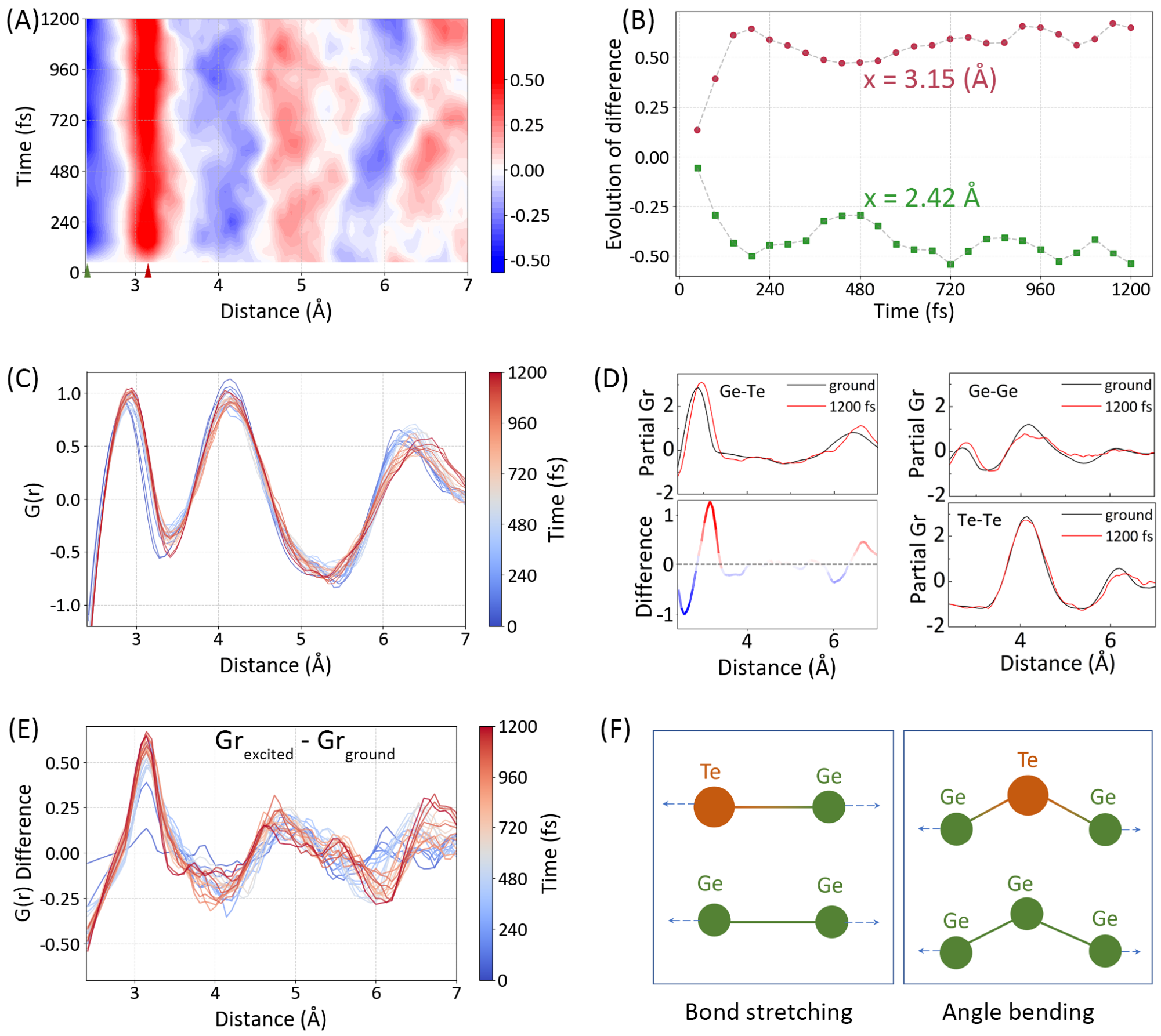

**Fig. 4. TDDFT-MD simulation of the ultrafast amorphous-amorphous transitions. (A)** Time-resolved ΔPDF in the false colour map with TDDFT-MD simulation. **(B)** Temporal evolution of the ΔPDF at two distinct distances. **(C)** Reduced pair distribution function G(r), i.e. PDF, as a function of the distance and the time delay. **(D)** Distribution of the Ge-Te, Ge-Ge and Te-Te distance of G(r) at the ground state and the time delay of 1200 fs. **(E)** ΔPDF as a function of the time delay and the distance. **(F)** Schematic presentation of the bond stretching and angle bending.

To further quantify the ultrafast bond stretching and angle bending dynamics after photoexcitation, we perform the time-dependent density-functional theory molecular dynamic simulation (TDDFT-MD, see details of the simulation method in Supplementary Materials). The simulation results with 3% electronic excitation are shown in Fig. 4. The time-resolved ΔPDF in Fig. 4A agrees well with the experimental results in Fig. 2C. In Fig. 4B, the ultrafast intensity change within 0.2 ps and the opposite changes at 2.42 Å and 3.15 Å aling with the experimental results in Fig. 2D. A week antiphase oscillation between the two curves can be identified with a frequency a bit larger than the experimental results in Fig. 2D. In Fig. 4C, the first peak of the PDF around 2.6 Å displays a clear ultrafast right shift, which coincides with the ultrafast bond stretching regime shwon in Fig. 2. By decomposing the PDF into atom-dependent distance distribution, Fig. 4D displays the Ge-Ge, Ge-Te and Te-Te distance at the ground state and the excited state at the time dealy of 1200 fs. Clearly, both the Ge-Te and Ge-Ge bond at ~ 2.6 Å display the bond stretching characteristic. Moreover, we further confirm the bond stretching regime by analyzing the characteristic distribution of atomic triplets (see Fig. S3 in Supplementary Materials). Another notable feature in Fig. 4D is the anti-correlated intensity variation of the Ge-Ge distance around the ~ 4.0 Å and the ~ 5.4 Å, which agrees well with the experimental results in Fig. 3 and evidences that the angle bending regime is governed by the Ge-Te (Ge)-Ge motif. In the time domain, the transition of ΔPDF from negative to positive at ~ 5.4 Å is clearly visible in Fig. 4A and Fig. 4E, consistent with the experimental observations presented in Fig. 3. The above simulation results exclusively evidence the ultrafast strectching of the Ge-Te (Ge) bond and the ultrafast angle bending of the Ge-Te (Ge)-Te motif, which reproduce the experimental results. The schematic illustration of the ultrafast bond stretchting and angle bending is shown in Fig. 4F.

Accompanying the ultrafast amorphous-amorphous transitions, the electronic structure undergoes a drastic transformation from an insulating state to a metallic state, as shown in Fig. 5 (3% electronic excitation) and Fig. S4 (1% electronic excitation) in Supplementary Materials. At the ground state (Fig. 5A), the system exhibits a disordered amorphous structure and a small but finite band gap near the Fermi level, characteristic of an insulator. Upon electronic excitation, at 1.2 ps (Fig. 5B), the structural disorder is preserved, confirming retention of the amorphous nature, while the band gap vanishes, signaling a transition to a metallic state. This metallic state emerges within hundreds of femtoseconds and persists for picoseconds. Notably, the excited electrons are recombined in the simulation of electronic structures, therefore, this insulator-metal transition is not sustained by persistent electronic excitation but is instead driven by the ultrafast amorphous-amorphous structural rearrangement. As illustrated in the central panel of Fig. 5, this structural change arises from collective atomic motions involving bond stretching and angle bending.

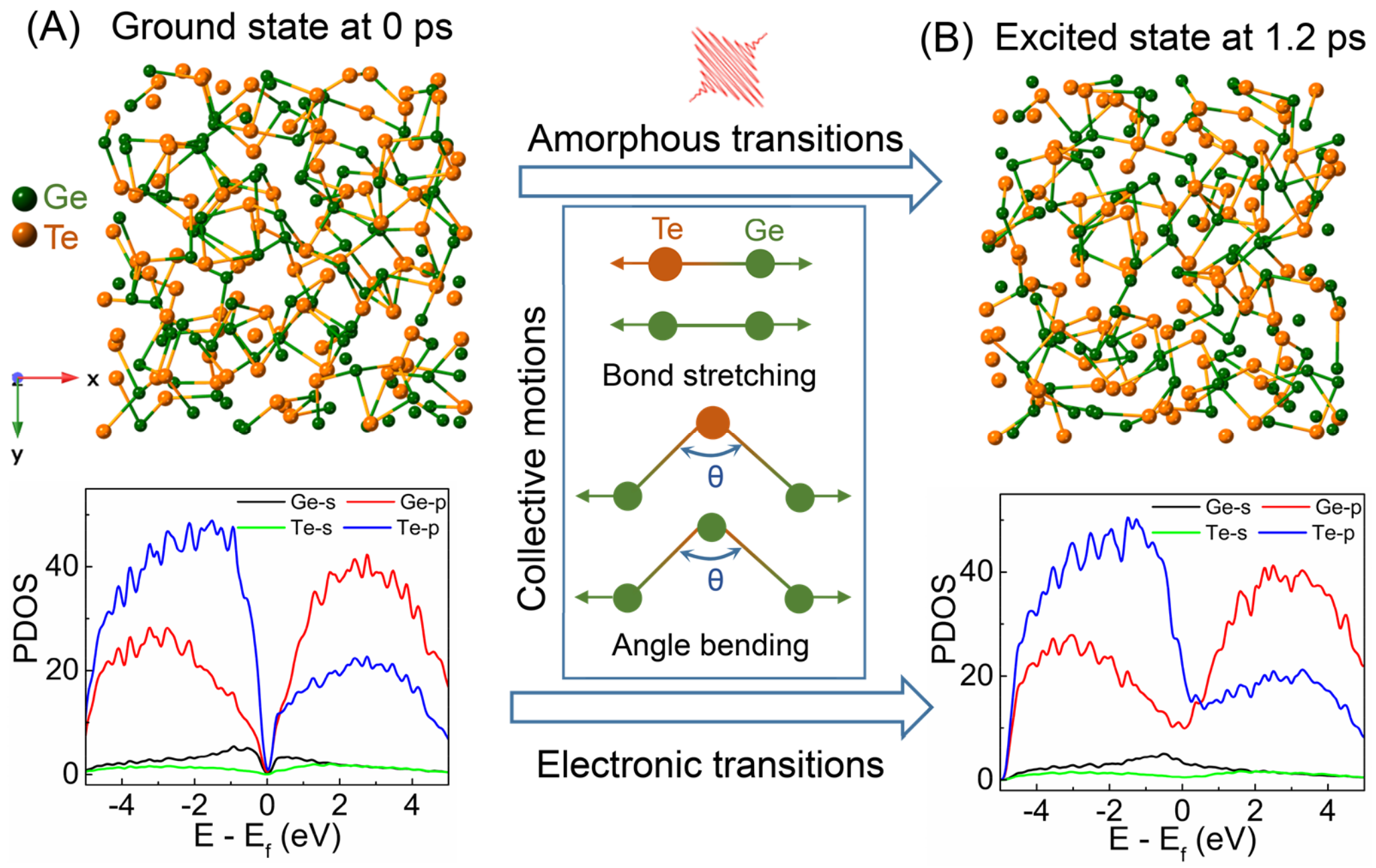

**Fig. 5. TDDFT-MD simulation of the ultrafast electronic transitions by collective atomic motions. (A)** Amorphous structure (top) and electronic structure (bottom) at the ground state. **(B)** Amorphous structure (top) and electronic structure (bottom) at the excited state (1.2 ps). The middle is the schematic of the collective atomic motions driving the amorphous transitions and the electronic transitions.

## Disscussion and conclusion

In this study, with femtosecond electron diffraction and state-of-the-art TDDFT-MD simulation, we identify the ultrafast amorphous-amorphous transitions indicated by ultrafast bond stretching and angle bending in amorphous GeTe. These ultrafast collective atomic motions from femtosecond to picosecond provide a clear strcutural origin in driving the ultrafast optical and electrical response in amorphous phase-change materials [69,70]. In previous study [27], the reinforcement of the local Peierls-like distortions by aging slows down the crystallization process. In our case, the ultrafast electronic delocalization by bond stretching and the ultrafast suppression of wrong bonds intrinsically reduce the stochastic crytal nucleation and imply an upper limit on the timescale for the formation of subcritical crystalline nuclei [69]. To advance the crystallization speed, a feasible solution is the double-pulse excitation regime, where one femtosecond laser/electric pulse excites firstly initiating the incubation process for nucleation and then, with a certain time delay, another picosecond pulse excites serving as a thermal bath for crystallization. Since the incubation process is excited collectively, the crystallization speed is expected to exceed the previously reported record of ~ 500 ps, which is achieved through random prestructural ordering [61,62]. Besides the photoexcited intermediate states, the intrinsic structural properties, such as the polarized bonding structure, the localized oscillator and the multiple local minimums, are revealed unambiguously with the femtosecond and picometer spatiotemporal resolution. Since the phonon picture is invalidated in glasses, these collective atomic motions excited by femtosecond laser are dominated by complex Coulomb interactions among electrons and nuclei. Moreover, these collective atomic motions construct the structure origin of the vibration modes of the boson peak in the THz range in real space. Taken together, our work demonstrates that femtosecond electron diffraction is a powerful in deciphering the structural complexity and functional mechanisms of amorphous materials-such as the instability of the local structure, the boson peak and the many-body correlation-by resolving collective atomic motions from random diffusion dynamics in the time domain. This capability provides critical experimental benchmarks for refining and validating theoretical models [21,42,43].

Ultrafast electronic excitation induced by femtosecond laser pulses serves as a close analogue to the electronic excitation triggered by high electric fields during ovonic threshold switching. Moreover, the ultrafast electronic excitation and the slower electric/thermal excitation in most cases trigger a universal transformation pathway [71,72]. Therefore, these collective atomic motions that drive the insulator-metal transition provide for the first time a comprehensive mechanistic understanding of the switching process and the speed limits for the ovonic threshold switching.

## Acknowledgements

**Funding:** This work was supported by National Natural Science Foundation of China (12304024, 12525501, 12335010) and National Key R&D Program of China (no. 2024YFA1612204). D.C. acknowledges the support from National Natural Science Foundation of China (Grant Numbers 12574348) and Natural Science Foundation of Anhui Province (Grant Numbers 2408085MF176). Work in Jilin University was supported by the National Natural Science Foundation of China (Grant No. 12274180) and the Science and Technology Development Plan Project of Changchun, China (Grant No. 2024GZZ07). The High-Performance Computing Center (HPCC) at Jilin University for computational resources is also acknowledged. D. X. would like to acknowledge the support from the New Cornerstone Science Foundation through the Xplorer Prize. The UED experiment was supported by the Shanghai soft x-ray free electron laser facility. **Author Contributions:** Yingpeng Qi devised the project and conceived the presented ideas. Min Zhu prepared and characterized the sample. Yingpeng Qi performed the MeV femtosecond electron diffraction experiments with help from Zhihui Zhou, Qing Xu, Yang Lv, Xiao Zou, Tao Jiang and Pengfei Zhu. Yingpeng Qi analyzed the experimental data and built up the model presented in the paper. Nianke Chen and Xianbin Li performed the TDDFT-MD simulation. Yingpeng Qi wrote the paper with contributions from all other authors. Yingpeng Qi, Nianke Chen, Dongxue Chen, Zhenrong Sun, Xianbin Li and Dao Xiang supervised the project. **Competing interests:** The authors declare no competing interest. **Data and materials availability:** All data needed to evaluate the conclusions in the paper are present in the paper and/or the Supplementary Materials. Additional data related to this paper may be requested from the corresponding author Yingpeng Qi.

**Supplementary Materials**

Materials and Methods

Fig. S1 to S3

References 67-80